\documentclass[aps,twocolumn,prd,showpacs,nofootinbib]{revtex4}
\usepackage{amsmath}
\usepackage{graphicx}
\usepackage{dcolumn}
\usepackage{bm}
\usepackage{amssymb}
\usepackage{latexsym}

\begin{document}

\title[Quantum States, Thermodynamic Limits and Entropy]
{Quantum States, Thermodynamic Limits and Entropy in M--Theory}
 
\author{M. C. B. Abdalla}
\address{Instituto de F\'{\i}sica Te\'orica, Universidade Estadual Paulista,
Rua Pamplona 145, 01405-900 - S\~ao Paulo, SP, Brazil\,\, 
{\em E-mail address:} {\rm mabdalla@ift.unesp.br}}

\author{A. A. Bytsenko}
\address{Departamento de F\'{\i}sica, Universidade Estadual de Londrina,
Caixa Postal 6001, Londrina-Paran\'a, Brazil \\
and \\
 Instituto de F\'{\i}sica Te\'orica, Universidade Estadual Paulista, 
Rua Pamplona 145, 01405-900 - S\~ao Paulo, SP, Brazil\,\, 
{\em E-mail address:} 
{\rm abyts@uel.br ; abyts@ift.unesp.br}}

\author{M. E. X.  Guimar\~aes}
\address{Departamento de 
Matem\'atica, Universidade de Bras\'{\i}lia, Bras\'{\i}lia, DF, Brazil \\
and \\
Instituto de F\'{\i}sica Te\'orica, Universidade Estadual Paulista,
Rua Pamplona 145, 01405-900 - S\~ao Paulo, SP, Brazil\,\, 
{\em E-mail address:} {\rm marg@unb.br ; emilia@ift.unesp.br}}

\date{September, 2003}

\thanks{}

\begin{abstract}

We discuss the matching of the BPS part of the spectrum for (super)membrane, 
which gives the possibility of getting membrane's results via string 
calculations. 
In the small coupling limit of M--theory the entropy of the 
system coincides with 
the standard entropy of type IIB string theory (including the 
logarithmic correction
term). The thermodynamic behavior at large 
coupling constant is computed by considering M--theory on a manifold with 
topology ${\mathbb T}^2\times{\mathbb R}^9$. We argue that the finite 
temperature partition functions (brane Laurent series for $p \neq 1$) 
associated with BPS $p-$brane spectrum can be analytically continued to  
well--defined functionals. It means that a finite temperature can be 
introduced in brane theory, which behaves like  finite temperature
field theory. In the limit $p \rightarrow 0$ (point particle limit) it gives 
rise to the standard behavior of thermodynamic quantities.

\end{abstract}

\maketitle

\section{Introduction}

There are deep connections between fundamental (super)membrane and 
(super)string theory. In particular, it has been shown that the BPS 
spectrum of states for type IIB string on a circle is in correspondence with 
the BPS spectrum of fundamental compactified supermembrane 
\cite{Schw,Russ}.
Brane thermodynamics can indicate non--trivial information about microscopic
degrees of freedom and the behavior of quantum systems at high temperature.  
Finite temperature M--theory defined on a manifold with topology 
${\mathbb T}^2\times{\mathbb R}^9$, at small and large string coupling 
constant regime, has been considered recently in \cite{Russo,Abdalla}. 
In the small radius of compactification limit M--theory recovers the  
type IIB superstring thermodynamics. In that case the critical temperature 
coincides with the Hagedorn temperature \cite{Russo}. There is a first 
order phase transition at temperature less than the Hagedorn temperature
 with a large 
latent heat leading to a gravitational instability \cite{Atick}. 

The purpose of the present paper is to consider the above mentioned 
problems, comparing small and large coupling regimes by considering 
M--theory on a 
manifold with topology ${\mathbb T}^2\times{\mathbb R}^9$, 
where one of the sides of the ${\mathbb T}^2$ torus is the Euclidean time 
direction (fermions obey antiperiodic boundary conditions). We turned to the 
problem of asymptotic density of quantum states for fundamental $p-$branes 
already initiated in \cite{Byts93,Byts94,Eliz94,Byts96}. 

This paper is organized as follows: In Section 2 the light--cone Hamiltonian 
formalism for membranes wrapped on a torus is summarized. The small coupling 
limit of M--theory
is considered in Section 3, while the limit of large coupling constant is 
analized in Section 4. We calculate the entropy associated with a string 
and argue that there is an interesting possibility allowing for a finite 
temperature being introduced into the brane theory. 
Section 5 summarizes our findings and discusses the relevant results.

\section{Toroidal membranes}

Let us consider the light--cone Hamiltonian formalism
for membranes wrapped on a torus in Minkowski space. A 
compactification of M--theory with (--,+) spin structure, having the topology
${\mathbb T}^2\times{\mathbb R}^9$, assumes that the dimensions $X^{11},\,
X^{10}$ are compactified on a torus with radii $R_{10},\, R_{11}$
and two spatial membrane directions wind around this torus.
The single--valued functions on the torus $X^{10}(\sigma,\rho),
X^{11}(\sigma,\rho)$, where $\sigma ,\rho \in [0,2\pi )$, are the membrane 
world--volume coordinates: 
\begin{eqnarray}
X^{10}(\sigma ,\rho )&=&m_0R_{10}\sigma + \widetilde X^{10} (\sigma ,\rho ),
\nonumber \\
X^{11} (\sigma ,\rho )&=&R_{11} \rho  + \widetilde X^{11} (\sigma ,\rho )
\mbox{.}
\end{eqnarray}
The eleven bosonic coordinates are
$\{ X^0, X^i,X^{10},X^{11}\}$ and the transverse coordinates 
$X^i(\sigma ,\rho )$, $i=1,2,...,8$ are all single--valued.
The transverse coordinates can be expanded in a complete basis 
of functions on the torus, namely
\begin{eqnarray}
X^i(\sigma,\rho )&=&\sqrt{\alpha '} \sum_{k,\ell} X^i_{(k,\ell)} e^{ik\sigma +
i\ell\rho },\nonumber \\
P^i(\sigma,\rho )&=&\frac{1}{(2\pi )^2 \sqrt{\alpha '}} \sum_{k,\ell} 
P^i_{(k,\ell)} 
e^{ik\sigma +i\ell\rho }
\mbox{.}
\end{eqnarray}
In these equations $\alpha'=\big( 4\pi ^2 R_{11} T_{2}\big)^{-1}$, while
$T_{2}$ is the membrane tension. 
The membrane Hamiltonian in light--cone formalism 
\cite{Berg,Wit,Russo2,Russo} is $H=H_0+H_{\rm int}$, 
where for bosonic modes of membrane the Hamiltonian takes the form: 
\begin{eqnarray}
\alpha'  H_0 & = & 8\pi^4 \alpha 'T_{2}^2 R_{10} ^2 R_{11}^2 m^2 
\nonumber \\
&  + & \frac{1}{2} 
\sum _{\bf n} 
\big[ P_{\bf n}^i P^i_{-{\bf n}}
+\omega_{km}^2 X^i_{\bf n} X^i_{-{\bf n} }\big]
\mbox{,}
\label{ham1}
\end{eqnarray}
\begin{equation}
\alpha'  H_{\rm int}= \frac{1}{4{\rm g}^2_A}\sum ({\bf n}_1 
\times {\bf n}_2)({\bf n}_3\times {\bf n}_4)
X_{{\bf n}_1}^i  X_{{\bf n}_2 }^j  X_{{\bf n}_3}^i   X_{{\bf n}_4}^j
\mbox{.}
\label{ham2}  
\end{equation}
In Eqs. (\ref{ham1}) and (\ref{ham2}) ${\bf n}\equiv (k,\ell)$, 
${\bf n}\times\ {\bf n}'=k\ell'-\ell k'$,
${\rm g}^2_A \equiv R_{11}^2(\alpha')^{-1}=4\pi^2R_{11}^3T_{2}$, 
$\omega_{k\ell} =(k^2 + m^2 \ell^2 R_{10}^2R_{11} ^{-2})^{1/2}$,
and $(m,k,\ell,k',\ell')\in {\mathbb Z}$.
The interaction term (\ref{ham2}) depends on the type IIA
string coupling ${\rm g}_A$. Mode operators, related to basic functions 
$X^{i}(\sigma,\rho),\,P^i(\sigma,\rho)$, are
\begin{eqnarray}
X^i_{(k,\ell)} & = & \frac{1}{i\sqrt {2} \omega_{(k,\ell)} }
\big[\alpha^i_{(k,\ell)}+\widetilde\alpha^i_{(-k,-\ell)}\big] \, , 
\nonumber \\
P^i_{(k,\ell)}& = & \frac{1}{\sqrt {2} }\big[\alpha^i_{(k,\ell)}-
\widetilde \alpha^i_{(-k,-\ell)}\big]
\mbox{,}
\end{eqnarray}
\begin{equation}
\big( X_{(k,\ell)}^i\big) ^\dagger =X_{(-k,-\ell)}^i\ ,\ \ \ \ 
\big( P_{(k,\ell)}^i\big) ^\dagger =P_{(-k,-\ell)}^i
\mbox{,}
\end{equation}
and $\omega_{(k,\ell)}\equiv{\rm sign}(k )\,\omega_{k\ell}$.
The canonical commutation relations read
\begin{eqnarray}
\big[ X^i_{(k,\ell)} , P^j_{(k',\ell')} \big] & = & i\delta_{k+k'}
\delta_{\ell+\ell'}\delta^{ij}\,, \nonumber  \\
\big[\alpha_{{(k,\ell)}}^i , \alpha^j_{(k',\ell')}\big] & = & 
\omega_{(k,\ell)} \delta_{k+k'}\delta_{\ell+\ell'}\delta^{ij}
\mbox{.}
\end{eqnarray}
The similar relations hold for the $\widetilde \alpha _{ {(k,\ell)} } ^i$.
The mass operator becomes 
\begin{equation}
M^2=2p^+p^--(p^{i})^2-p_{10}^2=2 (H_0+H_{\rm int})-(p^{i})^2-p_{10}^2
\mbox{.}
\end{equation}

The Hamiltonian of the membrane is non--linear, but there are two situations 
where one can simplify this Hamiltonian (we shall consider these cases
in the next sections):

\medskip
({\bf i}) {\em The limit ${\rm g}_A\rightarrow 0$.}

({\bf ii}) {\em The other limit of large ${\rm g}_A$.}

\section{Zero torus area limit of M--theory}

The zero torus area limit of M--theory on ${\mathbb T}^2$ leads to the
asymptotic ${\rm g}_A\rightarrow 0$ at fixed $(R_{10}/R_{11})$.
In M--theory it gives a ten--dimensional type IIB string.
More precisely, it has been shown \cite{Schw,Russo2} that quantum states 
of M--theory describe the $(p,q)$ strings bound states of type IIB 
superstring. 

Let us consider string theory in Euclidean space (time coordinate $X^0$ is 
compactified on a circle of circumference $\beta$). The presence of coordinates
compactified on circles gives rise to winding string states. The string 
single--valued function $X^0(\sigma,\tau )$ admits an expansion:
\begin{equation}
X^0(\sigma ,\tau )= x^0+ 2\alpha' p^0 \tau +2 R_0w_0\sigma + 
\widetilde X(\sigma,\tau )
\mbox{,}
\end{equation}
where $p^0=\ell_0(R_0)^{-1}$, \, $\ell_0,\,m_0 \in {\mathbb Z}$.
The Hamiltonian and the level matching constraints become 
\begin{eqnarray}
H & = & \alpha' p_i^2 +\frac{m_0^2R_0^2}{\alpha' }+ 
\frac{\alpha' \ell_0^2}{R^2_0}
+2(N_L+N_R-a_L-a_R)=0 \, ,  \nonumber \\
& &        N_L-N_R=\ell_0m_0 \, , 
\end{eqnarray} 
where $a_L, a_R$ are the normal ordering constants, which 
represent the vacuum energy of the (1+1)--dimensional field theory. 
In the case of type II superstring the number operators in the  
$m_0=\pm 1$ sector read 
\begin{eqnarray}
N_L & = & \sum_{{n}=1}^\infty \big[\alpha_{-{n}}^i \alpha _{n}^i +(n
-\frac{1}{2}) S_{-n}^a S^a_n \big]\ , \nonumber \\
 N_R & = & \sum_{n=1}^\infty \big[\widetilde \alpha_{-n}^i \widetilde 
\alpha _n^i +(n-\frac{1}{2})
\widetilde S_{-n}^a \widetilde S^a_n\big]\ ,
\end{eqnarray}
where $a=1,...,8$.
The normal--ordering constants are the same as in the NS sector of the NSR 
formulation, i.e. $a_L=a_R= 1/2$.

\subsection{The entropy in type II string theory}

To begin our discussion of the entropy in string theory 
we recall that the semiclassical quantization of $p-$branes, compactified 
on a manifold with topology 
${\frak M}= {\Bbb T}^p\times {\Bbb R}^{D-p}$, leads to the ``number operators''
$N_{\bf n}$ with ${\bf n}=(n_1,...,n_p)\in {\Bbb Z}^p$.
Therefore, let us consider multi--component versions of the classical 
generating functions for partition functions, namely
\begin{equation}
{\mathfrak G}_{\pm}(z)=\prod_{{\bf n}\in {\mathbb Z}^p/\{{\bf 0}\}}
\left[1\pm
\exp\left(-z\omega_{{\bf n}}({\bf a}, {\bf g})\right)\right]^{\pm \Lambda}
\mbox{,}
\end{equation}
where $\Re z>0$,\, $\Lambda>0$, $\omega_{{\bf n}}({\bf a}, 
{\bf g})$ is given by
$
\omega_{\bf n}({\bf a}, {\bf g})=
\left(\sum_{\ell}a_\ell(n_\ell+{\rm g}_\ell)^2\right)^{1/2}
$,
${\rm g}_\ell$, and $a_\ell$ are some real numbers.

In the context of thermodynamics of fundamental $p-$branes, classical 
generating functions ${\mathfrak G}_{\pm}(z)$ can be regarded as 
a partition function associated to fermi (or bose) modes, where 
$z\equiv\beta$ is the inverse temperature. In order to calculate the 
thermodynamic quantities we need first to know the total number of 
quantum states which can be described by the functions 
$\Omega_{\pm}(N)$ defined by
\begin{equation}
{\mathcal K}_{\pm}(t)=\sum_{N=0}^{\infty}\Omega_{\pm}(N)t^N 
\equiv {\mathfrak G}_{\pm}(-\log t)
\mbox{,}
\end{equation}
where $t<1$, and $N$ is a total quantum number. The Laurent inversion 
formula associated with the above definition has the form
\begin{equation}
\Omega_{\pm}(N)= \frac{1}{2\pi i}\oint dt\,t^{-N-1}{\mathcal K}_{\pm}(t)
\mbox{,}
\end{equation}
where the contour integral is taken on a small circle about the origin.
The $p-$dimensional Epstein zeta function 
$Z_p\left|_{\bf h}^{\bf g}\right|(z,\varphi)$ associated with the quadratic 
form $\varphi [{\bf a}({\bf n}+{\bf g})]=(\omega_{\bf n}({\bf a}, {\bf g}))^2$ 
for $\Re\,z>p$ is given by the formula
\begin{equation}
Z_p\left| \begin{array}{ll}
{\rm g}_1\,...\,{\rm g}_p \\
h_1\,...\,h_p\\
\end{array} \right|(z,\varphi)=\sum_{{\bf n}\in {\Bbb Z}^p}{}'
\left(\varphi[{\bf a}({\bf n}+{\bf g})]\right)^{-\frac{z}{2}}
e^{2\pi i({\bf n},{\bf h})}
\mbox{,}
\label{Epstein}
\end{equation}
where $({\bf n},{\bf h})=\sum_{i=1}^p
n_ih_i$,\, $h_i$ are real numbers and the prime on $\sum {'}$
means to omit the term ${\bf n} =-{\rm {\bf g}}$ if all the ${\rm g}_i$ are 
integers. For $\Re z<p$,\, $Z_p\left|_{\bf h}^{\bf g}\right|(z,\varphi)$ is
understood to be the analytic continuation of the right hand side of the
Eq. (\ref{Epstein}). The functional equation for 
$Z_p\left|_{\bf h}^{\bf g}\right|(z,\varphi)$ reads
\begin{eqnarray}
\!\!\!\!\!\!\!\!\!\!\!\!\!\!\!
Z_p\left| \begin{array}{ll}
{\bf g}\\
{\bf h}\\
\end{array} \right|(z,\varphi) & = & 
\frac{\pi^{\frac{1}{2}(2z-p)}}{({\rm det}\,{\bf a})^{1/2}}
\frac{\Gamma(\frac{p-z}{2})}
{\Gamma(\frac{z}{2})} \nonumber \\
& \times  & e^{-2\pi i({\bf g},{\bf h})}
Z_p\left| \begin{array}{ll}
{\bf\,\,\,\, h}\\
-{\bf g}\\
\end{array} \right|(p-z,\varphi^*)
\mbox{,}
\label{Epstein1}
\end{eqnarray}
and $\varphi^*[{\bf a}({\bf n}+ {\bf g})]=
\sum_\ell a_\ell^{-1}(n_\ell+{\rm g}_\ell)^2$.
Equation (\ref{Epstein1}) gives the analytic continuation of the zeta 
function. 
Note that $Z_p\left|_{\bf h}^{\bf g}\right|(z,\varphi)$ is an entire 
function in the complex $z-$plane except for the
case when all the $h_i$ are integers. In this case 
$Z_p\left|_{\bf h}^{\bf g}\right|(z,\varphi)$ has a simple pole
at $z=p$ with residue
$
A(p)= 2\pi^{p/2}[({\rm det}\,{\bf a})^{1/2}\Gamma(p/2)]^{-1}
$,
which does not depend on the winding numbers ${\rm g}_\ell$. 
Furthermore one has
$Z_p\left|_{\bf h}^{\bf g}\right|(0,\varphi)=-1$.

By means of the asymptotic expansion of 
${\mathcal K}_{\pm}(t)$ for $t\rightarrow 1$, 
which is equivalent to the ${\mathfrak G}_{\pm}(z)$ expansion for small 
$z$, one arrives at a complete asymptotic limit of $\Omega_{\pm}(N)$
\cite{Byts93,Byts94,Eliz94,Byts96}:
\begin{eqnarray}
 & & \Omega_{\pm}(N)_{N\rightarrow \infty}  =
{\mathcal C}_{\pm}(p)N^{\frac{2\Lambda Z_p\left|_{{\bf 0}}^{{\bf g}}\right|
(0,\varphi)-p-2}{2(1+p)}} \nonumber \\
& & \times  \exp\left\{\frac{1+p}{p}[\Lambda A(p)\Gamma(1+p)\zeta_{\pm}(1+p)]^
{\frac{1}{1+p}}N^{\frac{p}{1+p}}\right\} \nonumber \\
& & \times [1+{\mathcal O}(N^{-\kappa_{\pm}})]
\mbox{,}
\label{asym1}
\end{eqnarray}
\begin{eqnarray}
& & {\mathcal C}_{\pm}(p)=[\Lambda A(p)\Gamma(1+p)
\zeta_{\pm}(1+p)]^{\frac{1-2\Lambda 
Z_p\left|_{{\bf 0}}^{{\bf g}}\right|(0,\varphi)}{2p+2}} \nonumber \\
& & \times
\frac{\exp\left[\Lambda (d/dz)
Z_p\left|_{{\bf 0}}^{{\bf g}}\right|(z,\varphi)|_{(z=0)}\right]}
{[2\pi(1+p)]^{1/2}}
\mbox{,}
\label{asym2}
\end{eqnarray}
where $\zeta_{-}(z)\equiv \zeta_R(z)$ is the Riemann zeta function,
$\zeta_{+}(z) = (1-2^{1-z})\zeta_R(z)$,
$
\kappa_{\pm}= p/(1+p)\min \left({\mathcal C}_{\pm}(p)/p-\delta/4,
1/2-\delta\right)
$, and $0<\delta<2/3$.
Using  Eqs. (\ref{asym1}) and (\ref{asym2}) and assuming linear Regge 
trajectories, i.e. the mass formula $M^2=N$ for the number of brane states of 
mass $M$ to $M+dM$, one can obtain the asymptotic density for (super)$p-$brane 
states.

In fact, for linear Regge--like trajectories the partition function always
diverges. This IR divergence in the partition function might be regularized by 
some effects of brane theory, for example, like imposing U--duality 
(see, for example, \cite{Bytsenko}) or choosing non--linear behavior of 
Regge trajectory (say, $M^{(1+p)/p}$ or something similar).

These results can be used in the context of the brane method's calculation of
the ground state degeneracy of systems with quantum numbers of certain BPS
extreme black holes \cite{Callan,Maldacena,Vafa,Halyo}. 
The brane picture gives the entropy in
terms of partition functions ${\mathfrak G}_{\pm}(z)$ for a gas of 
species of massless quanta. In fact for unitary conformal theories of fixed 
central charge $c$ Eq. (\ref{asym1}) represents the degeneracy of the 
state $\Omega (N)$ with momentum $N$ and for $N\rightarrow \infty $ one 
has \cite{Abdalla,Abdalla1}: 
\begin{equation}
{\mathcal S}(N)={\rm log} \Omega (N)\simeq {\mathcal S}_{0} +
{\mathcal A}(p,c){\rm log} ({\mathcal S}_{0}) 
\mbox{,}
\end{equation}
where ${\mathcal S}_{0}= {\mathcal A}_0\sqrt{cN}$ and 
${\mathcal A}_0$ is a real number. It gives the growth of the degeneracy 
of BPS solitons for $N \gg 1$. Note that in the case of zero modes the 
dependence of the logarithmic correction ${\mathcal A}(p,c)$ on an embedding 
spacetime can be eliminate \cite{Rama}.

\section{Large string coupling limit}

We now focus on the case ({\rm {\bf ii}}) mentioned at the end of Section 2 
by letting the constant ${\rm g}_A$ being large. In this limit 
$R_{10},\,R_{11}$ 
are large with fixed
$(R_{10}/R_{11})$ and the non--linear interacting Hamiltonian is multiplied 
by the small constant ${\rm g}_A^{-2}$ so that it can be considered 
perturbatively. In the leading order of perturbative theory in 
${\rm g}_A^{-2}$ the interaction term can be dropped and the solution of
the membrane equations of motion takes the form \cite{Russo}
\begin{eqnarray}
\!\!\!\!\!\!\!\!\!\!\!\!\!\!\!\!\!\!\!\!\!\!\!\!\!\!\!\!\!\!
& & X^i (\sigma, \rho, \tau )= x^i +\alpha'  p^i \tau  \nonumber \\
\!\!\!\!\!\!\!\!\!\!\!\!\!\!\!\!
& & + 
\sqrt{-\frac{\alpha'}{2}} \sum_{{\bf n} \neq (0,0)}
\frac{e^{i w_{\bf n} \tau }}{\omega_{\bf n}}
\big[ \alpha _{\bf n}^i e^{ik\sigma +i\ell\rho } 
+ \widetilde \alpha _{\bf n} ^i e^{-ik\sigma -i\ell\rho }\big] 
\mbox{.}
\end{eqnarray}
The momentum components in the $X^{10}$ and $X^{11}$ directions 
are: 
$p_{10}=(\ell_{10}/R_{10}),$ and $p_{11}=(\ell_{11}/R_{11})$,
where $\ell_{10},\,\ell_{11}\in {\mathbb Z}$. The nine--dimensional mass 
operator reads
\begin{eqnarray}
\!\!\!\!\!\!\!\!\!\!\!\!\!\!\!\!\!\!
M^2  & = & \frac{\ell^2_{10}}{R_{10}^2} + \frac{\ell^2_{11}}{R_{11}^2} + 
\frac{m_0^2 R_{10}^2}{ 
\alpha  ^{\prime 2}}\ \nonumber \\
\!\!\!\!\!\!\!\!\!\!\!\!\!\!
& + &   \frac{1}{\alpha' }
\sum _{k,\ell} \big( \alpha^i_{(-k,-\ell)} \alpha^i_{ (k,\ell)} 
+ \widetilde \alpha^i_{(-k,-\ell)} \widetilde \alpha^i_{(k,\ell)}\big)
\mbox{.}
\end{eqnarray}
The level--matching conditions are \cite{Duff,Russo}:
$
N_\sigma^+ - N_\sigma^- = m_0 \ell_{10},\,\,\, 
N_\rho^+ - N_\rho ^- = \ell_{11}
$, and
\begin{eqnarray}
N^+_\sigma & = &  \sum _{\ell=-\infty }^\infty \sum _{k=1}^\infty 
\frac{k}{\omega_{k\ell} }
\alpha^i_{(-k,-\ell)} \alpha^i_{(k,\ell)}
\mbox{,} \nonumber \\
N^-_\sigma & = &  \sum _{\ell=-\infty }^\infty \sum _{k=1}^\infty \frac{k}
{\omega_{k\ell} }
\widetilde \alpha^i_{(-k,-\ell)} \widetilde \alpha^i_{(k,\ell)}
\mbox{,}
\end{eqnarray}
\[
N^+_\rho  =  \sum_{\ell=1}^\infty \sum_{k=0}^\infty 
\frac{\ell}{\omega_{k\ell} }
\big[ \alpha^i_{(-k,-\ell)} \alpha^i_{(k,\ell)} + 
\widetilde \alpha^i_{(-k,\ell)} \widetilde \alpha^i_{(k,-\ell)} \big]
\mbox{,} 
\]
\begin{equation}
N^-_\rho  = \sum_{\ell=1}^\infty \sum_{k=0}^\infty 
\frac{\ell}{\omega_{k\ell} }
\big[ \alpha^i_{(-k,\ell)}\alpha^i_{(k,-\ell)} + 
\widetilde \alpha^i_{(-k,-\ell)} \widetilde \alpha^i_{(k,\ell)} \big]
\mbox{.}
\end{equation}
Let us define the quantum oscillator operator ${\widehat H}$ as 
\begin{equation}
{\widehat H}= \sum _{k,\ell} \big(: \alpha^i_{(-k,-\ell)} 
\alpha^i_{ (k,\ell)} : + :\widetilde \alpha^i_{(-k,-\ell)} 
\widetilde \alpha^i_{ (k,\ell) }:\big)
\mbox{,}
\label{H}
\end{equation}
where the annihilation operators $\alpha _{ {(k,\ell)} }^i, \ 
\widetilde \alpha _{ {(k,\ell)}}^i$ are determined for $k>0$ and 
$\ell\in {\mathbb Z}$, and $k=0$, $\ell>0$. In Eq. (\ref{H}) the normal 
ordering means taking the annihilation operators to the right. 
The relation is (see \cite{Russo}):
$
H=  {\widehat H}+ 2(D-3)E,\,\,\,
E= (1/2) \sum_{k,\ell} \omega_{k\ell}
$,
where the constant energy shift $2(D-3)E$\,\,\,($E$ is the Casimir energy) 
represents the purely bosonic contribution to the vacuum energy of the 
(2+1)--dimensional field theory. In the case of supersymmetry preserving 
boundary conditions for fermions the contributions to the vacuum energy 
coming from bosonic and fermionic fields cancel out \cite{Berg1,Duff}. 
\\

In the presence of membrane excitation states with non--trivial
winding numbers around the target space torus the spectrum of the 
light--cone Hamiltonian is discrete \cite{Berg1,Duff,Russo}. Let the Euclidean 
time coordinate $X^0$ play the role of $X^{10}$. Then fermions will obey 
antiperiodic boundary conditions around $X^0$ but periodic boundary conditions 
around $X^{11}$. In the  $m_0=\pm 1$ sector fermions are antiperiodic under 
the replacement $\sigma\rightarrow\sigma+2\pi$ (while periodic under 
$\rho\rightarrow\rho+2\pi$). The Hamiltonian operator becomes
\begin{equation}
{\mathcal H}= \frac{\ell^2_{0}}{R_{0}^2} + \frac{\ell^2_{11}}{R_{11}^2} + 
\frac{R_{0}^2}{\alpha  ^{\prime 2}}\  + \frac{1}{\alpha' } ({\widehat H} + 
2(D-3)E)
\mbox{,}
\end{equation}
where 
\begin{eqnarray}
{\widehat H} & = & \sum _{\bf n} \big[ :\alpha^i_{-{\bf n}} \alpha^i_{{\bf n}} 
:+ :\widetilde \alpha^i_{-{\bf n}} \widetilde \alpha^i_{{\bf n}}: 
\nonumber \\
& + &
\omega_{k+ \frac{1}{2} ,\ell} \big( :S^a_{-{\bf n}} S^a_{{\bf n}} : 
+ :\widetilde S^a_{-{\bf n}} \widetilde S^a_{{\bf n}}:\big)\big]
\mbox{,}
\end{eqnarray}
and 
\begin{eqnarray}
&& E = E_{\rm B}+ E_{\rm F}= \frac{1}{2} 
\sum_{k,\ell} \big( \omega_{k\ell}- \omega_{k+\frac{1}{2}, \ell} \big)\,,
\nonumber \\
&& \omega_{k\ell} =  \bigg({k^2} +\frac{\ell^2}{{\rm g}_{eff} ^2} \bigg)^{1/2}
\mbox{.}
\end{eqnarray}

\subsection{Brane thermodynamics presented in M--theory}

Let us consider semiclassically the partition function associated with 
fundamental $p-$branes (which is known to be divergent) embedded in flat 
$D-$dimensional manifolds.  
For the standard quantum field model the free energy associated with bosonic 
(b) and fermionic (f) degrees of freedom has the form (see, for example,
\cite{Eliz94,Byts96})
\begin{eqnarray}
F^{(b,f)}(\beta) & = &
- \pi^p({\rm det}{\mathfrak A})^{1/2}
\int_0^{\infty}\frac{ds\,\Xi^{(b,f)}(s,\beta)}{(2s)^{(D-p+2)/2}}
\nonumber \\
& \times &
\Theta\left[\begin{array}{r}
{\bf g}\\
{\bf 0}
\end{array}\right]({\bf 0}|\Omega)
e^{-sM_0^2/2\pi}
\mbox{,}
\label{free}
\end{eqnarray}
where
\begin{eqnarray}
\Xi^{(b)}(s,\beta) & = & \theta_3\left(0\Big\vert\frac{i\beta^2}{2s}\right)-1\,,
\nonumber \\
\Xi^{(f)}(s,\beta) & = & 1-\theta_4\left(0\Big\vert\frac{i\beta^2}{2s}
\right)
\mbox{,}
\end{eqnarray}
and $\theta_3(\nu|\tau)$ and $\theta_4(\nu|\tau)=
\theta_3(\nu+\frac{1}{2}|\tau)$ are the Jacobi theta functions.
Here ${\mathfrak A}={\rm diag}(R_1^{-2},...,R_p^{-2})$ is a 
$p\times p$ matrix. The
global parameters $R_\ell$ characterizing the non--trivial topology  
appear in the theory due to the fact that the coordinates 
$x_\ell (\ell=1,...,p)$ obey
the conditions $0\leq x_\ell<2\pi R_\ell$. The number of topological 
configurations
of quantum fields is equal to the number of elements in group 
$H^1({\mathfrak M}; {\mathbb Z}_2)$, that is, the first cohomology group 
with coefficients in ${\mathbb Z}_2$. 
The multiplet ${\bf g}=({\rm g}_1,...,{\rm g}_p)$ defines the topological type
of field (i.e., the corresponding twist), and depends on the field type
chosen in ${\mathfrak M}$, ${\rm g}_\ell=0$ or $1/2$. In our case 
$H^1({\mathfrak M};{\mathbb Z}_2)={\mathbb Z}_2^p$ and so the number of 
topological configurations of real scalars (spinors) is $2^p$. 

We follow the notations and treatment of 
\cite{Mumf84} and introduce the theta function with characteristics
${\bf a}, {\bf b}$ for ${\bf a},{\bf b}\in{\mathbb Z}^p$,
\begin{equation}
\Theta\left[\begin{array}{r}
{\bf a}\\
{\bf b}
\end{array}\right]({\bf z}|\Omega) = \sum_{{\bf n}\in {\Bbb Z}^p}
e^{
\pi i\left[({\bf n}+{\bf a})\Omega({\bf n}+{\bf a})
+ 2({\bf n}+{\bf a})({\bf z}+ {\bf b})\right]}\,.
\end{equation}
In this connection $\Omega=(s i /2\pi^2)\mbox{diag}(R_1^2,...,R_p^2)$. 

We assume that the free energy is equivalent to a sum of
the free energies of quantum fields which are present in the modes of a 
$p-$brane.
The factor $\exp(-sM_0^2/2\pi)$ in Eq. (\ref{free}) should be understood as
$\mbox{Tr}\exp(-sM^2/2\pi)$, where $M$ is the mass operator of the brane 
and the trace is taken over an infinite set of Bose--Fermi oscillators 
$N_{{\bf n}}$. The one--loop--like contribution for the (super)$p-$brane
can be evaluated making use of the Mellin--Barnes representation 
for the partition function (energy integral) and in this formalism the 
generating function reads 
\begin{equation*}
{\mathfrak G}_{\pm}(z) =
{\rm Tr}\left[e^{-zM^2}\right]_{\pm} = 
\frac{1}{2\pi i}\int_{\Re\,s=s_0}\!\!\!
ds\Gamma(s){\rm Tr}[zM^2]^{-s}_{\pm}\,,
\end{equation*}
\begin{equation}
{\rm Tr}[zM^2]^{-s}_{\pm}=
\frac{1}{\Gamma(s)}
\int_0^\infty dtt^{s-1}
{\mathfrak G}_{\pm}(tz)
\mbox{.}
\end{equation}
One can use some substraction procedure for the divergent terms in
${\mathfrak G}_{\pm}(z)$ in order to procede with the regularization 
scheme 
(see for detail Ref. \cite{Byts1}). To simplify our calculation we set
$a_\ell=R_\ell=1$, and the final result for the free energy is 
\cite{Actor,Byts2}:
\begin{eqnarray}
F_{\pm}(\beta) & \simeq & -Q(D,p)
\sum_{k=1}
^{\infty}\frac{\Gamma\left(pk+\frac{1-p}{2}\right)}{\Gamma(k)}
\nonumber \\
& \times &
\zeta_{\pm}(2pk+1-p)
x^{1+p(2k-1)} 
\mbox{,}
\label{serie}
\end{eqnarray}
where $Q(D,p)=\Lambda A(p)\Gamma(p)[y(p)]^{-1-p}$, with
 $\Lambda =D-p-1$\,,
and
\begin{equation}
x \beta = y(p) \equiv 
\left[\Lambda 2^{3p-2}\pi^{\frac{p-1}{2}}\Gamma\left(\frac{p+1}{2}\right)
\zeta_R(p+1)\right]^{\frac{1}{2p}}
\mbox{.}
\end{equation}
The asymptotic expansion of $\Gamma(s)$ for large value of $|s|$ has the form
\begin{equation}
\Gamma(s) = (2\pi)^{\frac{1}{2}}s^{s-\frac{1}{2}}e^{-s}
\left(1+{\mathcal O}(s^{-1})\right),
\,\,\,\,\,\,\,\,|{\rm arg}\,s|<\pi
\mbox{,}
\end{equation}
and for $p>1$ the power series (\ref{serie}) is divergent for any $x>0$.

\subsection{The analytic continuation of a brane  Laurent series and the 
thermodynamic limit}

In principle, the power series (\ref{serie}) is divergent, nevertheless one 
can 
construct its analytic continuation. Let us define for 
$|z|<\infty$ two series
\begin{equation}
{\mathfrak W}_{\pm}(z)=\sum_{k=0}^{\infty}\frac{\sqrt{\pi}\nu_{\pm}(k;p)}
{\Gamma(k+1)
\Gamma\left(pk+\frac{p+2}{2}\right)}\left(\frac{z}{2}\right)^
{p(2k+1)+1}
\mbox{,}
\end{equation}
where the factors $\nu_{\pm}(k;p)$ have the form
\begin{eqnarray}
\nu_{-}(k;p) & = & (-1)^{pk+1}\,, 
\nonumber \\
\nu_{+}(k;p) & = & \nu_{-}(k;p)\left[1-2^{-p(2k+1)-1}\right]
\mbox{.}
\end{eqnarray}
For finite variable $z$ these series converge and the convergence improves  
rapidly with the increasing of the integer number $p$. Let 
$z=\ell\cdot2\pi x$, then we get the series
\begin{equation}
\sum_{\ell=1}^{\infty}{\mathfrak W}_{\pm}(\ell\cdot2\pi x)
=\sum_{\ell=1}^{\infty}
\sum_{k=0}^{\infty}\frac{\sqrt{\pi}\nu_{\pm}(k;p)(\ell\pi x)^{p(2k+1)+1}}
{\Gamma(k+1)\Gamma\left(pk+\frac{p+2}{2}
\right)}
\mbox{.}
\label{sum}
\end{equation}
Now, if we commute the (up to now divergent) sum $\Sigma_\ell$ with the sum 
$\Sigma_k$, new 
extra terms of the type $x^{-1}W_{\pm}(p)$ will appear on the right hand 
side of Eq. (\ref{sum}). Therefore, the result is
\begin{eqnarray}
&&
\!\!\!\!\!\!\!\!\!\!\!\!\!\!
\sum_{\ell=1}^{\infty}{\mathfrak W}_{\pm}(\ell2\pi x) + x^{-1}W_{\pm}(p)
\nonumber \\
&& 
\!\!\!\!\!\!\!\!\!\!\!\!\!\!
= \sum_{k=0}^{\infty}\frac{\pi\nu_{\pm}(k;p)}{\Gamma(k+1)\Gamma\left(
pk+\frac{p+2}{2}\right)}\frac{\zeta_{R}[-p(2k+1)]}
{x^{-1-p(2k+1)}}
\nonumber \\
&& 
\!\!\!\!\!\!\!\!\!\!\!\!\!\!
= \sin\left(\frac{\pi p}{2}\right)\sum_{k=1}^{\infty}
\frac{\Gamma\left(pk+\frac{1-p}{2}\right)}{\Gamma(k)}
\frac{\zeta_{\pm}(2pk+1-p)}
{x^{-1-p(2k-1)}}
\mbox{,}
\label{3line}
\end{eqnarray}
where $W_{\pm}(p)$ is an integer function of $p$ (see, for example, 
\cite{Actor}). In the second equality the functional equation for 
$\zeta_{R}(s)$ has been  used. 

The new form of $F({\beta})$ is: 
\begin{eqnarray}
F_{-}(\beta)& \simeq & \frac{Q(D,p)}{\sin\left(\frac{\pi p}{2}\right)}      
\sum_{k=0}^{\infty}\frac{(-1)^{pk}\pi\zeta_{R}[-p(2k+1)]}
{\Gamma(k+1)\Gamma\left(pk+\frac{p+2}{2}\right)}
\nonumber \\
& \times &
\left(\frac{\beta}{y(p)}\right)^{-1-p(2k+1)}
\mbox{.}
\label{F}
\end{eqnarray}

In the string case ($p=1$) the corresponding series in Eq. (\ref{serie}) 
can be resummed into the trigonometrical form using the identities
\begin{eqnarray}
\sum_{k=1}^{\infty}\zeta_{-}(2k)x^{2k} & = & \frac{1}{2}\left(1- 
\pi x {\rm cot}(\pi x)\right)\,,
\nonumber \\
\sum_{k=1}^{\infty}\zeta_{+}(2k)x^{2k} & = &
\frac{\pi x}{4}{\rm tan}\left(\frac{\pi x}{2}\right)
\mbox{.}
\label{trig}
\end{eqnarray}
The finite radius of convergence, $\mid x \mid < 1$, of the Laurent series 
 corresponds to the
Hagedorn temperature in string thermodynamics (see for detail 
Ref. \cite{Byts1}). Using trigonometric relations, formulae (\ref{trig})
display a certain periodicity in the temperature. The physical meaning of that
behaviour is still obscure. The thermal dependence in Eqs. (\ref{F}), 
(\ref{trig}), corresponding to the quantum modes in two dimensions near the 
Hagedorn
instability, can be interpreted as an indication of a vast reduction of the 
fundamental degrees of freedom in string theory \cite{Atick}.
   
The divergent series in Eq. (\ref{serie}) for the case $p>1$, when 
reexpressed on the left hand side of Eq. (\ref{3line}), remains  
well--defined for finite temperature. Note that the series (\ref{F}) has 
smooth $\beta\rightarrow 0, \infty$ limits. For example, when
$\beta\rightarrow \infty$ we get:
\begin{equation}
F_{-}(\beta) \simeq {\mathcal A}_1(D,p)\beta^{-1-p} + {\mathcal A}_2(D,p)
\beta^ {-1-3p} + {\mathcal O}(\beta^{-1-5p}) 
\mbox{,}
\label{final}
\end{equation} 
where ${\mathcal A}_{\ell}(D,p)\,(\ell = 1,2)$ depends on the dimension of the 
embedding spacetime. The statistical internal energy 
$E= (\partial/\partial\,\beta)(\beta F_{-}(\beta))$ and the entropy 
$S = \beta^ 2 (\partial/\partial\,\beta)F_{-}(\beta)$ can be easily
 calculated 
using Eq. (\ref{F}). The $\beta-$behavior has a similar dependence which is 
similar to the one found for the string case in \cite{Atick, Bytsenko95}.
Note that in the  $p \rightarrow 0$ limit (point particle limit) Eq. (\ref{F})
leads to the standard thermodynamic behavior  for both $F,E \sim T$.

\section{Conclusions}

The result (\ref{asym1}), (\ref{asym2}) has an universal character. We can 
compute state density, including the prefactors ${\mathcal C}_{\pm}(p)$, 
depending on the dimension of the embedding space. 
There are deep connections between strings and $p-$branes; at least they 
should be considered as different limits of a more general M--theory. Indeed, 
string results may be obtained via membrane--string correspondence and vice 
versa. Therefore, even being not a fundamental theory of $p-$branes it may 
provide new deep insights in the understanding of string theory and consistent 
formulation of M--theory.

In this paper we had dealt with the same discrete membrane spectrum 
as it has been used in the membrane--string correspondence. We analyzed
the logarithmic correction to the entropy in the small string coupling 
limit of M--theory. 
Note that in the large string coupling limit of M--theory, compactified on 
manifold with topology ${\mathbb T}^2\otimes{\mathbb R}^9$, 
the analytic continuation of a brane Laurent series has been given in its 
explicit form. Physically, it means that a finite temperature can be 
introduced in the 
theory and a membrane (if it can be quantized semi--classically) behaves like 
an ideal gas of quantum modes, which corresponds to a field theory at finite 
temperature (zero critical temperature). Finally note that in the limit 
$p \rightarrow 0$ (point particle limit) the standard behavior of 
thermodynamic quantities has been obtained. 
\\
\\

\section*{Acknowledgements}

A.A.B. and M.E.X.G. would like to thank Funda\c{c}\~ao de Amparo 
\`a Pesquisa do Estado de S\~ao Paulo (FAPESP/Brazil) for 
partial support 
and the Instituto de F\'{\i}sica Te\'orica (IFT/UNESP) for kind hospitality. 
M.C.B.A. and A.A.B. would like to thank the Conselho Nacional de 
Desenvolvimento Cient\'{\i}fico e Tecnol\'ogico (CNPq/Brazil) for 
partial support.

\end{document}